\documentclass[aps,prb,twocolumn,showpacs,superscriptaddress]{revtex4}
\usepackage{amsmath}
\usepackage{amssymb}
\usepackage{epsfig}
\usepackage{color}
\usepackage{graphicx,amsmath}

\begin{document}

\title{Quasi-classical physics and $\bf \it T$-linear
resistivity in both strongly correlated and ordinary metals}
\author{V. R. Shaginyan}\email{vrshag@thd.pnpi.spb.ru}
\affiliation{Petersburg Nuclear Physics Institute, Gatchina,
188300, Russia}\affiliation{Clark Atlanta University, Atlanta, GA
30314, USA} \author{K. G. Popov}\affiliation{Komi Science Center,
Ural Division, RAS, Syktyvkar, 167982, Russia}
\author{V. A. Khodel} \affiliation{Russian Research Centre Kurchatov Institute,
Moscow, 123182, Russia} \affiliation{McDonnell Center for the Space
Sciences \& Department of Physics, Washington University,
St.~Louis, MO 63130, USA}

\begin{abstract}

We show that near a quantum critical point generating quantum
criticality of strongly correlated metals where the density of
electron states diverges, the quasi-classical physics remains
applicable to the description of the resistivity $\rho$ of strongly
correlated metals due to the presence of a transverse zero-sound
collective mode, reminiscent of the phonon mode in solids. We
demonstrate that at $T$, being in excess of an extremely low Debye
temperature $T_D$, the resistivity $\rho(T)$ changes linearly with
$T$, since the mechanism, forming the $T$ dependence of $\rho(T)$,
is the same as the electron-phonon mechanism that prevails at high
temperatures in ordinary metals. Thus, in the region of the
$T$-linear resistivity, electron-phonon scattering leads to near
material-independence of the lifetime $\tau$ of quasiparticles that
is expressed as the ratio of the Planck constant $\hbar$ to the
Boltzmann constant $k_B$, $T\tau\sim \hbar/k_B$. We find that at
$T<T_D$ there exists  a different mechanism, maintaining the
$T$-linear dependence of $\rho(T)$, and making the constancy of
$\tau$ fail in spite of the presence of $T$-linear dependence. Our
results are in good agreement with exciting experimental
observations.

\end{abstract}


\maketitle

Discoveries of surprising universality in the properties of both
strongly correlated metals and ordinary ones provide unique
opportunities for checking and expanding our understanding of
quantum criticality in strongly correlated compounds. When
exploring at different temperatures $T$ a linear in temperature
resistivity of these utterly different metals, an universality of
their fundamental physical properties has been revealed.
\cite{bruin} On one hand, at low $T$ the linear $T$-resistivity
\begin{equation}
\rho(T)=\rho_0+AT,\label{res}
\end{equation}
observed in many strongly correlated compounds such as
high-temperature superconductors and heavy-fermion metals located
near their quantum critical points and therefore exhibiting quantum
criticality. Here $\rho_0$ is the residual resistivity and $A$ is a
$T$-independent coefficient. Explanations based on quantum
criticality for the $T$-linear resistivity have been given in the
literature, see e.g. \cite{varma,varma1,phill,phill1} and Ref.
therein. On the other hand, at room temperatures the $T$-linear
resistivity is exhibited by conventional metals such as $\rm Al$,
$\rm Ag$ or $\rm Cu$. In case of a simple metal with a single Fermi
surface pocket the resistivity reads $e^2n\rho=p_F/(\tau v_F)$,
\cite{trio} where $e$ is the electronic charge, $\tau$ is the
lifetime, $n$ is the carrier concentration, and $p_F$ and $v_F$ are
the Fermi momentum and the Fermi velocity, correspondingly. Writing
the lifetime $\tau$ (or inverse scattering rate) of quasiparticles
in the form \cite{tomph,arch}
\begin{equation}\label{LT}
\frac{\hbar}{\tau}\simeq a_1+\frac{k_BT}{a_2},
\end{equation}
we obtain
\begin{equation}\label{vf}
a_2\frac{e^2n\hbar}{p_Fk_B}\frac{\partial\rho}{\partial
T}=\frac{1}{v_F},
\end{equation} where $\hbar$ is Planck's constant, $k_B$ is
Boltzmann's constant, $a_1$ and $a_2$ are $T$-independent
parameters. A challenging point for a theory is that experimental
facts corroborate Eq. \eqref{vf} in case of both strongly
correlated metals and ordinary ones provided that these demonstrate
the linear $T$-dependence of their resistivity. \cite{bruin}
Moreover, the analysis of data available in literature for the most
various compounds with the linear dependence of $\rho(T)$ shows:
The coefficient $a_2$ is always close to unit, $0.7\leq a_2\leq
2.7$, notwithstanding huge distinction in the absolute value of
$\rho$, $T$ and Fermi velocities $v_F$, varying by two orders of
magnitude. \cite{bruin} As a result, it follows from Eq. \eqref{LT}
that the $T$-linear scattering rate is of universal form, $1/(\tau
T)\sim k_B/\hbar$, regardless of different systems displaying the
$T$-linear dependence. Indeed, this dependence is demonstrated by
ordinary metals at temperatures higher than the Debye temperature,
$T\geq T_D$, with an electron - phonon mechanism and by strongly
correlated metals which are assumed to be fundamentally different
from the ordinary ones, in which the linear dependence at their
quantum criticality and temperatures of a few Kelvin is assumed to
come from excitations of electronic origin rather than from
phonons. \cite{bruin} We note that in some of the cuprates the
scattering rate has a momentum and doping dependence omitted in Eq.
\eqref{vf}. \cite{peets,french,alld} Nonetheless, the fundamental
picture outlined by Eq. \eqref{vf} is strongly supported by
measurements of the resistivity on $\rm Sr_3Ru_2O_7$ for wide range
of temperatures: At $T\geq 100$ K, the resistivity becomes again
the $T$-linear at all applied magnetic fields, as it does at low
temperatures and at the critical field $B_{c}\simeq 7.9$ T, but
with the coefficient $A$ lower than that seen at low temperatures.
\cite{bruin} Thus, the same strongly correlated compound exhibits
the same behavior of the resistivity at both quantum critical
regime and high temperature one, allowing us to expect that the
same physics governs the $T$-linear resistivity in spite of
possible peculiarities of some compounds.

In this paper we show that the same physics describes the
$T$-linear dependence of the resistivity of both conventional
metals and strongly correlated metals at their quantum criticality.
As an example, we analyze the resistivity of $\rm Sr_3Ru_2O_7$, and
demonstrate that our results are in good agreement with
experimental facts.

\begin{figure} [! ht]
\begin{center}
\includegraphics [width=0.47\textwidth]{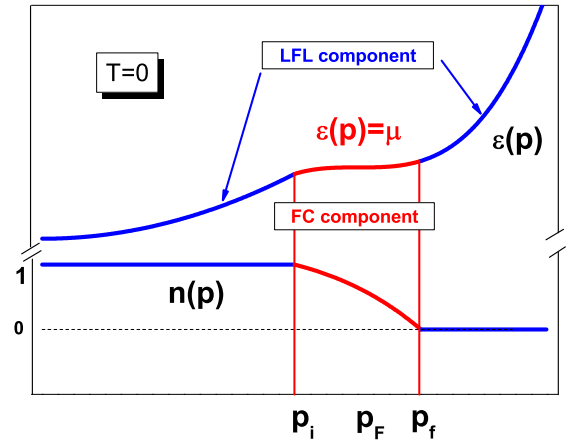}
\end{center}
\caption {(Color on line) Schematic plot of two-component electron
liquid at $T=0$ with FC. Due to the presence of FC, the system is
separated into two components: The first component is a normal
liquid with the quasiparticle distribution function $n_0(p<p_i)=1$,
and $n_0(p>p_f)=0$; The second one is FC with $0<n_0(p_i<p<p_f)<1$
and the single-particle spectrum $\varepsilon (p_i<p<p_f)=\mu$. The
Fermi momentum $p_F$ satisfies the condition $p_i<p_F<p_f$.}
\label{fig1}
\end{figure}

To develop explanations of constancy of $T$-linear scattering rate
$1/(\tau T)$, it is necessary to recall the nature and consequences
of flattening of single-particle excitation spectra
$\varepsilon({\bf p})$ (``flat bands'') in strongly correlated
Fermi systems \cite{khod,khs,volovik,noz} (for recent reviews, see
\cite{shagrep,shag,mig100}). At $T=0$, the ground state of a system
with a flat band is degenerate, and the occupation numbers
$n_0({\bf p})$ of single-particle states belonging to the flat band
are continuous functions of momentum ${\bf p}$, in contrast to
discrete standard Landau Fermi liquid (LFL) values 0 and 1, as it
seen from Fig. \ref{fig1}. Such behavior of $n_0({\bf p})$ leads to
a temperature-independent entropy term
\begin{equation}
S_0=-\sum_{\,{\bf p}} [n_0({\bf p})\ln n_0({\bf p})+(1-n_0({\bf
p}))\ln(1- n_0({\bf p}))]. \label{S0}
\end{equation}
Unlike the corresponding LFL entropy, which vanishes linearly as
$T\to 0$, the term $S_0$ produces the non-Fermi liquid (NFL)
behavior that includes $T$-independent thermal expansion
coefficient. \cite{arch,shagrep,aplh,plas} That $T$-independent
behavior is observed in measurements on $\rm CeCoIn_5$
\cite{oes,steg,zaum} and $\rm YbRh_2(Si_{0.95}Ge_{0.05})_2$,
\cite{kuch} while very recent measurements on $\rm Sr_3Ru_2O_7$
indicate the same behavior. \cite{gegprl,gegen} In the theory of
fermion condensation (FC), the degeneracy of the NFL ground state
is removed at any finite temperature, with the flat band acquiring
a small dispersion \cite{noz,shagrep}
\begin{equation}
\varepsilon({\bf p})=\mu+T\ln \frac{1-n_0({\bf p})}{n_0({\bf p})}
\label{tem}
\end{equation}
proportional to $T$ with $\mu$ being the chemical potential. The
occupation numbers $n_0$ of FC remain unchanged at relatively low
temperatures and, accordingly, so does the entropy $S_0$. Due to
the fundamental difference between the FC single-particle spectrum
and that of the remainder of the Fermi liquid, a system having FC
is, in fact, a two-component system. The range $L$ of momentum
space adjacent to the Fermi surface where FC resides is given by
$L\simeq (p_f-p_i)$, as seen from Fig. \ref{fig1}.

In strongly correlated metals at high temperatures, a light
electronic band coexists with $f$ or $d$-electron narrow bands
placed below the Fermi surface. At lower temperatures when the
quantum criticality is formed, a hybridization between this light
band and $f$ or $d$-electron bands results in its splitting into
new flat bands, while some of the bands remain light representing
LFL states. \cite{merge} A flat band can also be formed by a van
Hove singularity (vHs).
\cite{sc4,sc7,sc9,ragt,pnas,sigr,berr,physc} We assume that at
least one of these flat bands crosses the Fermi level and
represents the FC subsystem shown in Fig. \ref{fig1}. Remarkably,
the FC subsystem possesses its own set of zero-sound modes. The
mode of interest for our analysis is that of transverse zero sound
with its $T$-dependent sound velocity $c_t\simeq\sqrt{T/M}$ and the
Debye temperature \cite{DP}
\begin{equation} T_D\simeq c_tk_{max}\simeq \beta\sqrt{TT_F}.\label{td}
\end{equation}
Here, $\beta$ is a factor, $M$ is the effective mass of electron
formed by vHs or by the hybridization, $T_F$ is the Fermi
temperature, while $M^*$ is the effective mass formed finally by
some interaction, e.g. the Coulomb interaction, generating flat
bands. \cite{merge} The characteristic wave number $k_{max}$ of the
soft transverse zero-sound mode is estimated as $k_{max}\sim p_F$
since we assume that the main contribution forming the flat band
comes from vHs or from the hybridization. We note that the
numerical factor $\beta$ cannot be established and is considered as
a fitting parameter, rendering of $T_D$ given by Eq. \eqref{td}
correspondingly uncertain. Estimating $T_F\sim 10$ K and taking
$\beta\sim 0.3$, and observing that the quasi-classical regime
takes place at $T>T_D\simeq \beta\sqrt{TT_F}$, we obtain that
$T_D\sim 1$ K and expect that strongly correlated Fermi systems can
exhibit a quasi-classical behavior at their quantum criticality
\cite{DP,kpla} with the low-temperature coefficient $A$ entering
Eq. \eqref{res} $A=A_{LT}$. In case of HF metals with their few
bands crossing Fermi level and populated by LFL quasiparticles and
by HF quasiparticles, the transverse zero sound make the
resistivity possesses the $T$-linear dependence at the quantum
criticality as the normal sound (or phonons) do in the case of
ordinary metals. \cite{kpla} It is quite natural to assume that the
sound scattering in these materials is near material-independent,
so that electron-phonon processes both in the low temperature limit
at the quantum criticality and in the high temperature limit of
ordinary metals have the same $T$-linear scattering rate that can
be expressed as
\begin{equation}\label{tau}
\frac{1}{\tau T}\sim \frac{k_B}{\hbar}.
\end{equation}
Thus, in case of the same material the coefficient $A=A_{HT}$,
defining the classical linear $T$-dependence generated by the
common sound (or phonons) at high temperatures, coincides with that
of low-temperature coefficient $A_{LT}$, $A_{HT}\simeq A_{LT}$. As
we shall see, this observation is in accordance with measurements
on $\rm Sr_3Ru_2O_7$. \cite{bruin} It is worth noting that the
transverse zero sound contribution to the heat capacity $C$ follows
the Dulong-Petit law, making $C$ possess a $T$-independent term
$C_0$ at $T\gtrsim T_D$, as it does in case of ordinary metals.
\cite{DP} It is obvious that the zero sound contributes to the heat
transport as the normal sound does in case of ordinary metals, and
its presence can violate the Wiedemann-Franz low; a detailed
consideration of the emergence of zero sound and its properties
will be published elsewhere.

There is another mechanism contributing to the $T$-linear
dependence at the quantum criticality that we name the second
mechanism in contrast to the first one described above and related
to the transverse zero sound. We turn to consideration of the next
contribution to the resistivity $\rho$ in the range of quantum
criticality, at which the dispersion of the flat band is governed
by Eq. \eqref{tem}. It follows from Eq. \eqref{tem} that the
temperature dependence of $M^*(T)$ of the FC quasiparticles is
given by
\begin{equation} M^*(T)\sim
\frac{\eta p_F^2}{4T}, \label{M*}
\end{equation}
where $\eta=L/p_F$. \cite{shagrep,shag,mig100} Thus, the effective
mass of FC quasiparticles diverges at low temperatures, while their
group velocity, and hence their current, vanishes and the main
contribution to the resistivity is provided by light quasiparticles
bands.  Nonetheless, the FC quasiparticles still play a key role in
determining the behavior of both the $T$-dependent resistivity and
$\rho_0$. The resistivity has the conventional dependence
\cite{trio}
\begin{equation}\label{rhoo}
\rho(T)\propto M^*_{L}\gamma
\end{equation}
on the effective mass and damping of the normal quasiparticles.
Based on a fact that all the quasiparticles have the same lifetime,
one can show that in playing its key role, the FC makes all
quasiparticles of light and flat bands possess the same unique
width $\gamma$ and lifetime $\tau_q$ given by Eq. \eqref{LT}.
\cite{arch,jetpl} As a result, the first term $a_0$ on the right
hand side of Eq. \eqref{LT} forms an irregular residual resistivity
$\rho_0^c$, while the second one forms the $T$-dependent part of
the resistivity. The term ``residual resistivity'' ordinarily
refers to impurity scattering. In the present case, the irregular
residual resistivity $\rho_{0}^c$ is instead determined by the
onset of a flat band, and has no relation to scattering of
quasiparticles by impurities. \cite{arch} The two mechanisms
described above contribute to the coefficient $A$ on the right hand
side of Eq. \eqref{res} and it can be represented  as $A\simeq
A_{LT}+A_{FC}$, where $A_{LT}$ and $A_{FC}$ are formed by the zero
sound and by FC, respectively. Coefficients $A_{LT}$ and $A_{FC}$
can be identified and differentiated experimentally, for $A_{LT}$
is accompanied by the temperature independent heat capacity $C_0$,
while $A_{FC}$ is escorted by the emergence of $\rho_0^c$.

A few comments are in order here. As we have seen above, the
presence of flat bands generates the characteristic behavior of the
resistivity. Besides, it has a strongly influence on the systems
properties by creating the term $S_0$, making the spin
susceptibility of these systems exhibit the Curie-Weiss law, as it
is observed in the HF metal $\rm CeCoIn_5$. \cite{khod} The term
$S_0$ serves as a stimulator of phase transitions that could lift
the degeneracy and make $S_0$ vanish in accordance with the Nernst
theorem. As we shall see, in case of $\rm Sr_3Ru_2O_7$ the nematic
transition emerges. If a flat band is absent, the $T$-dependence of
the resistivity is defined by the dependence of the term $\gamma$,
entering Eq. \eqref{rhoo}, on the effective mass $M^*(T)$ of heavy
electrons, while the spin susceptibility is determined by $M^*(T)$.
\cite{shagrep}

We now consider the HF compound $\rm Sr_3Ru_2O_7$ to illustrate the
emergence of the both mechanisms contributing to the linear
$T$-dependence of the resistivity. To achieve a connected picture
of the quantum critical regime underlying the the quasi-classical
region in $\rm Sr_3Ru_2O_7$, we have to construct its $T-B$ phase
diagram. We employ the model
\cite{sc4,sc7,sc9,ragt,pnas,sigr,berr,physc} based on vHs that
induces a peak in the single-particle density of states (DOS) and
leads a field-induced flat band. \cite{flat} At fields in the range
$B_{c1}<B<B_{c2}$, the vHs is moved through the Fermi energy and
the DOS peak turns out to be at or near the Fermi energy. A key
point in this scenario is that within the range $B_{c1}<B<B_{c2}$,
a repulsive interaction (e.g., Coulomb) is sufficient to induce FC
and formation of a flat band with the corresponding DOS singularity
locked to the Fermi energy. \cite{shagrep,shag,mig100,flat} Now, it
is seen from Eq. \eqref{tem} that finite temperatures, while
removing the degeneracy of the FC spectrum, do not change the
excess entropy $S_0$, threatening the violation of the Nernst
theorem.  To avoid such an entropic singularity, the FC state must
be altered as $T\to 0$, so that $S_0$ is to be removed before zero
temperature is reached. This can take place by means of some phase
transition or crossover, whose explicit consideration is beyond the
scope of this paper. In case of $\rm Sr_3Ru_2O_7$, this mechanism
is naturally identified with the electronic nematic transition.
\cite{sc4,sc7,sc9}

\begin{figure}[!ht]
\begin{center}
\includegraphics [width=0.47\textwidth]{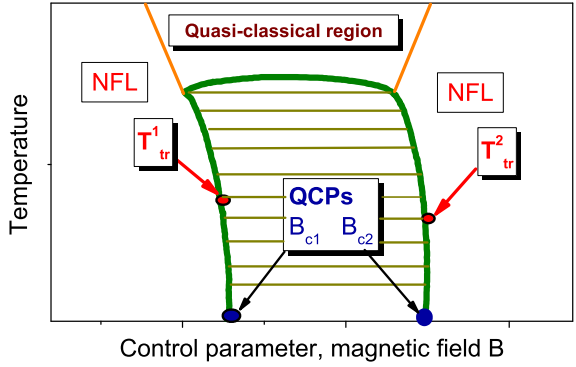}
\end{center}
\caption{(Color on line) Schematic phase diagram of the metal $\rm
Sr_3Ru_2O_7$. The quantum critical points (QCPs) situated at the
critical magnetic fields $B_{c1}$ and $B_{c2}$ are indicated by
arrows. The ordered phase bounded by the thick curve and demarcated
by horizontal lines emerges to remove the entropy excess given by
Eq. \eqref{S0}. Two arrows label the tricritical points $T^1_{\rm
tr}$ and $T^2_{\rm tr}$ at which the lines of second-order phase
transitions change to the first order. Quasi-classical region is
confined by two lines at the top of the figure and by the top line
of the ordered phase.} \label{fig2}
\end{figure}
The schematic $T-B$ phase diagram of $\rm Sr_3Ru_2O_7$ based on the
proposed scenario is presented in Fig.~\ref{fig2}.  Its main
feature is the magnetic field-induced quantum critical domain
created by quantum critical points situated at $B_{c1}$ and
$B_{c2}$, generating FC and associated flat band.  In contrast to
the typical phase diagram of a HF metal, \cite{shagrep} the domain
occupied by the ordered phase in Fig.~\ref{fig2} is seen to be
approximately symmetric with respect to the magnetic field
$B_c\simeq (B_{c2}+B_{c1})/2\simeq 7.9$ T. \cite{pnas} The emergent
FC and quantum critical points are considered to be hidden or
concealed in a phase transition. The area occupied by this phase
transition is indicated by horizontal lines and restricted by the
thick boundary lines. At the critical temperature $T_c$ where the
new (ordered) phase sets in, the entropy is a continuous function.
Therefore the top of the domain occupied by the new phase is a line
of second-order phase transitions. As $T$ is lowered, some
temperatures $T^1_{\rm tr}$ and $T^2_{\rm tr}$ are reached at which
the entropy of the ordered phase becomes larger than that of the
adjacent disordered phase, due to the remnant entropy $S_0$ from
the highly entropic flat-band state. Therefore, under the influence
of the magnetic field, the system undergoes a first-order phase
transition upon crossing a sidewall boundary at $T=T^1_{\rm tr}$ or
$T=T^2_{\rm tr}$, since entropy cannot be equalized there. It
follows, then, that the line of second-order phase transitions is
changed to lines of first-order transitions at tricritical points
indicated by arrows in Fig.~\ref{fig2}.  It is seen from
Fig.~\ref{fig2} that the sidewall boundary lines are not strictly
vertical, due to the stated behavior of the entropy at the boundary
and as a consequence of the magnetic Clausius-Clapeyron relation
(as discussed in \cite{sc9,ragt}). Quasi-classical region is
located above the top of the second order phase transition and
restricted by two lines shown in Fig. \ref{fig2}. Therefore, the
$T$-linear dependence is located in the same region and represented
by $AT$ with $A\simeq A_{LT}+A_{FC}$. We predict that in this
region the heat capacity $C$ contains the temperature independent
term $C_0$ as that of the HF metal $\rm YbRh_2Si_2$ does,
\cite{wolf} while jumps of the residual resistivity, represented by
$\rho_0^c$ in $\rm Sr_3Ru_2O_7$, \cite{sc4} are generated by the
second mechanism. \cite{arch,flat}

\begin{figure}[!ht]
\begin{center}
\includegraphics [width=0.52\textwidth]{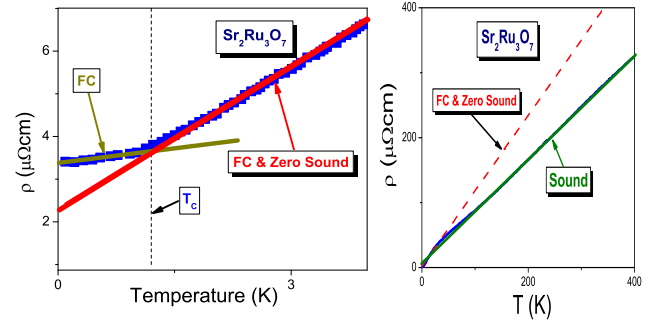}
\end{center}
\caption{(Color on line) The left panel: The resistivity $\rho(T)$
for $\rm Sr_3Ru_2O_7$ at the critical field of $B_c=7.9$ T.
\cite{pnas} Two straight lines display the $T$-linear dependence of
the resistivity exhibiting a kink at $T=T_c$. At $T>T_c$ the
$T$-linear resistivity is formed by the zero sound and FC
contributions, while at $T<T_c$ the linear part of the resistivity
comes from the FC contribution. The right panel: The resistivity at
$B_c$ over an extended temperature range up to 400 K. \cite{bruin}
The dash line shows the extrapolation of the low-$T$-linear
resistivity at $T>T_c$, and the solid line shows the extrapolation
of the high-$T$-linear resistivity formed at $T>100$ K by the
common sound.} \label{fig3}
\end{figure}
The coefficients $A_{FC}$, $A_{LT}$ and $A_{HF}$ can be extracted
from measurements of the resistivity $\rho(T)$ shown in the left
and right panels of Fig. \ref{fig3}. \cite{bruin,pnas} For the sake
of clearness, the left panel shows only a part of the data on
$\rho(T)$ that was measured from $0.1$ K to $18$ K at $B_c$, and
exhibits the $T$-linear dependence between $1.4$ K and $18$ K and
between $0.1$ K and $1$ K. \cite{pnas} The coefficient $A\simeq
A_{LT}+A_{FC}\simeq 1.1$ $\rm \mu\Omega cm/K$  between $18$ K and
$1.4$ K. Since $T_D\sim 1$ K, we expect that between $1$ K and
$0.1$ K the coefficient $A$ is formed by the second mechanism and
$A_{FC}\simeq 0.25$ $\rm \mu\Omega cm/K$. The right panel reports
the measurements of $\rho(T)$ for $T>T_c$ up to $400$ K.
\cite{bruin} The dash line shows the extrapolation of the
low-temperature linear resistivity at $T<20$ K and $B_c$ with
$A\simeq 1.1$ $\rm \mu\Omega cm/K$, and the solid line shows the
extrapolation of the high-temperature linear resistivity at $T>100$
K with $A_{HT}\simeq 0.8$ $\rm \mu\Omega cm/K$. \cite{bruin} The
obtained values of $A$ allow us to estimate the coefficients
$A_{LT}$ and $A_{FC}$. Due to our assumption that $A_{LT}\simeq
A_{HT}$, we have $A-A_{LH}\simeq A_{FC}\simeq 0.3$ $\rm \mu\Omega
cm/K$, this value is in good agreement with $A_{FC}\simeq 0.25$
$\rm \mu\Omega cm/K$. As a result, we conclude that for $\rm
Sr_3Ru_2O_7$ with its precise measurements the scattering rate is
given by Eq. \eqref{tau}, and does not depend on $T$, provided that
$T\geq T_D$ and the relatively small term $A_{FC}$ is omitted. On
the other hand, at $T<T_D$ $A_{HT}/A_{FC}\simeq 3$ and the
constancy of the lifetime $\tau$ is violated, while the resistivity
exhibits the $T$-linear dependence. It is seen from the left panel
of Fig. \ref{fig3}, that the change from the resistivity
characterized by the coefficient $A_{LT}$ to the resistivity with
$A_{FC}$ is seen as a kink at $T_c=1.2$ K representing both the
entry into the ordered phase and a transition region at which the
resistivity alters it slope. We expect that the constancy can also
fail in such HF metals as $\rm YbRh_2Si_2$ and the quasicrystal
$\rm Au_{51}Al_{34}Yb_{15}$ that exhibits the heavy-fermion
behavior. \cite{QCM,QCT}

\end{document}